\magnification=1200
\advance\hoffset by -.6truecm

\font\hd=cmbx10 scaled\magstep1
\font\draftfont=cmti7
\fontdimen16\tensy=2.7pt
\fontdimen17\tensy=2.7pt
\fontdimen14\tensy=5pt
\def\pagenumbers{\pageno=1\footline={\hss\tenrm\folio\hss}}
\def\today{\ifcase\month
\or January\or February\or March\or April\or May
\or June\or July\or August\or September\or October
\or November \or December\fi \space\number\day, \number\year}
\def\draft{\headline={\draftfont \the\pageno
\hfill File:\jobname, Draft Version:\today}}
\countdef\refno=30
\refno=0
\countdef\sectno=31
\sectno=0
\countdef\chapno=32
\chapno=0
\countdef\equationno=33
\equationno=0
\def\ref{\advance \refno by 1 \ifnum\refno<10 \item{ [\the\refno ]} \else
\item{[\the\refno ]} \fi}
\outer\def\section#1\par
{\global\advance\sectno by 1
\vskip0pt plus .3\vsize\penalty-250\vskip 0pt plus-.3\vsize
\noindent\hangit{\rlap{\bf\the\sectno.}\phantom{AAA}}{\kern-0.5em\bf#1}

\nobreak\smallskip}
 
\def\abstract{\vfill\eject{\bf Abstract}\smallskip}

\def\adno{\global\advance\equationno by 1}
\def\feqno{\adno\eqno(\the\sectno.\the\equationno)}
\def\hangit#1#2\par{\setbox1=\hbox{#1\enspace}
\hangindent\wd1\hangafter=0\noindent\hskip-\wd1
\hbox{#1\enspace}\ignorespaces#2\par}
\newdimen\hanglength
\hanglength=0truecm
\def\p_hangit#1{\ifdim\hanglength=0truecm
\hanglength=3truecm\fi\hangit{\rlap{#1}\hbox to\hanglength{\hfill}}}

\def\signature #1#2
{\par\vbox{\leftskip3in\parskip0pt\parindent0pt
\vskip1cm
\nobreak\penalty50
#1\par
#2}}
\def\sigle{\headline{\ifnum \pageno=1
\hfill {\hd DEPARTEMENT DE PHYSIQUE THEORIQUE}
\else\hfill\fi}
\includegraphics{//delos/user/mala/psfiles/sigle.ps}}
\def\integer{{Z \kern -.45 em Z}}
\def\complex{{\kern .1em {\raise .47ex \hbox {$\scriptscriptstyle |$}}
\kern -.4em {\rm C}}}
\def\real{{\vrule height 1.6ex width 0.05em depth 0ex
\kern -0.06em {\rm R}}}
\def\rational{{\kern .1em {\raise .47ex \hbox{$\scripscriptstyle |$}}
\kern -.35em {\rm Q}}}
\def\natural{{\vrule height 1.6ex width .05em depth 0ex \kern -.35em {\rm N}}}
\def\vide{{{\rm O} \kern -0.7em /}}

\parskip 0.5truecm
\baselineskip=12pt
\catcode`\^^?=9
\newskip\zmineskip \zmineskip=0pt plus0pt minus0pt
\mathchardef\mineMM=20000
\newinsert\footins
\def\footnote#1{\let\minesf\empty 
  \ifhmode\edef\minesf{\spacefactor\the\spacefactor}\/\fi
   #1\minesf\vfootnote{#1}}
\def\vfootnote#1{\insert\footins\bgroup
  \interlinepenalty\interfootnotelinepenalty
  \splittopskip\ht\strutbox 
  \splitmaxdepth\dp\strutbox \floatingpenalty\mineMM
  \leftskip\zmineskip \rightskip\zmineskip
\spaceskip\zmineskip \xspaceskip\zmineskip
 \item{#1}\footstrut\futurelet\next\fominet}
\def\fominet{\ifcat\bgroup\noexpand\next \let\next\fmineminet
  \else\let\next\fminet\fi \next}
\def\fmineminet{\bgroup\aftergroup\minefoot\let\next}
\def\fminet#1{#1\minefoot}
\def\minefoot{\strut\egroup}
\def\footstrut{\vbox to\splittopskip{}}
\skip\footins=\bigskipamount 
\count\footins=1000 
\dimen\footins=8in 
\newbox\fnsintbox
\newdimen\fnsintdimen
\def\fondsnational{\hfuzz=15pt\parindent=0pt\nopagenumbers%
\setbox\fnsintbox=\hbox{\bf 2.4.5 \enspace\hfill}%
\fnsintdimen=\wd\fnsintbox
\leftskip=\fnsintdimen}
\def\back#1 {\hskip-\leftskip\rlap{#1}\hskip\leftskip\hskip-.1em}
\def\bback#1 #2 {\hskip-\leftskip\rlap{\bf#1}\hskip\leftskip{\bf#2}}
\def\grsim{\kern2pt\hbox{\raise1pt\hbox{$>$}%
           \kern-8pt\lower4pt\hbox{$\sim$}}\kern2pt}
\def\smsim{\kern2pt\hbox{\raise1pt\hbox{$<$}%
           \kern-8pt\lower4pt\hbox{$\sim$}}\kern2pt}
 
\def\ref{\advance \refno by 1 \ifnum\refno<10 \item{ [\the\refno ]} \else
\item{[\the\refno ]} \fi}
\def\real{{\vrule height 1.6ex width 0.05em depth 0ex
\kern -0.06em {\rm R}}}
\def\smalreal{{\vrule height 1.1ex width 0.05em depth 0ex
\kern -0.06em {\rm R}}}
 
\def\pagenumbers{\pageno=1\footline={\hss\tenrm\folio\hss}}
\nopagenumbers
\hbox{}\vskip 6truecm
\centerline{{\bf CONTROVERSY ABOUT QUANTUM TIME DECAY}}
\vskip 2truecm
\centerline{W.O. Amrein}
\centerline{D\'epartement de Physique Th\'eorique}
\centerline{Universit\'e de Gen\`eve}
\centerline{CH-1211 Gen\`eve 4, Switzerland}
\vskip 4truecm
\centerline{UGVA-DPT 2001/03-1094}
\vfill\eject
\pagenumbers
\noindent
In the recent physics literature there have appeared contradictory statements
concerning the behaviour of scattering solutions of the 3-dimensional Schr\"odinger
equation at large times.
We clarify the situation and point out that the issue was rigorously resolved
in the mathematics literature.
\vskip1cm

In their papers [1] and [2], Garc\'\i a-Calder\'on, Mateos and Moshinsky considered
the nonescape probability $P(t)$ to find a quantum particle, subject to a potential
of finite range, still confined in some given finite spatial region at a large
time $t$. They claim that $P(t) \sim t^{-1}$ as $t \to \infty$.
On the other hand Muga, Delgado and Snider in [3] and Cavalcanti in a Comment
on [1] (see [4]) obtain a decay of $P(t)$ as $t^{-3}$.
In their Reply [5] to this Comment the authors of [1] reassert that
$P(t) \sim t^{-1}$.

The authors cited above seem to be unaware of the fact that, about two decades
ago, the question of time decay of solutions of the Schr\"odinger equation
was studied in considerable detail by mathematicians (some relevant
references are [6,7] for the 3-dimensional case and [8] for the
n-dimensional case, $n = 1,2,3,\dots$).
The results of these papers clearly imply that $P(t) \sim t^{-3}$ (see e.g.
\S 4 of [6], Theorems 10.3-10.5 in [7] or Theorem 7.6 and Example 7.8 in [8]).
More precisely, for a large class of potentials in 1 or
3 dimensions, one has $P(t) = O(t^{-3} )$ except possibly for a discrete set of
values of the coupling constant (essentially, these exceptional values of the
coupling constant are those for which there is a zero energy resonance,
and in such a situation the asymptotics of $P(t)$ may be only $O(t^{-1} )$).
For generic values of the coupling constant the $t^{-3}$ decay of $P(t)$ holds
for any bounded spatial domain and any wave packet $\Psi$ such that $\Psi (x,0)$
(the initial wave function at time $t = 0$) has no bound state components
and decays sufficiently rapidly at large $\vert x \vert$ (for example, if $n = 3$,
such that $\int_{\smalreal^3} (1+ \vert x \vert )^{2s} \vert \Psi (x,0)\vert^2 d^3
x < \infty$ for some $s > 5/2$ ).
For various initial states $\Psi$ the decay of $P(t)$ will be faster than
$t^{-3}$ (several terms in Eq. (7.25) of [8], when applied to $\Psi$, may be zero).

In the case of $s$-wave states as considered in [1], the nonescape
probability for a ball $B_R$ of radius $R$ (i.e. $B_R = \{ x \in \real^3\big\vert\vert x\vert
\equiv r \leq R\}$ ) is given as follows (see Eqs. (2) and (3) of [1])~:
$$
P(t) = \int^R_0 \Psi^*(r,t)\Psi (r,t)dr =
\int^R_0 dr \int^R_0 dr'g^* (r,r';t)\Psi^* (r',0)
\int^R_0 dr'' g(r,r'';t) \Psi (r'',0) \ ,
$$
where $g(r,r';t)$ denotes the time-dependent retarded Green's function.
To calculate $P(t)$ the authors of [1] express the Green's function
$g(r,r';t)$ as an infinite sum of resonant state functions and freely
interchange the order of integrations and summations.
In the absence of a justification for these interchanges, the expression (14)
for $P(t)$ given in [1] (identical with (2.41) in [2]) can not be considered to
be exact (in [1] and in the Reply [5], the incorrect time decay $P(t) \sim t^{-1}$
is obtained as a consequence of Eq. (14)).
It should also be mentioned that all calculations have to be done for a
{\it fixed} (possibly large) value of the time parameter $t$;
the argumentation presented in [5] - invoking an interchange of the long time
limit and the integration over the variable $r$ - is beside the point.

The author thanks J.G. Muga for drawing his attention to the disagreement on time
decay and for discussions.

\noindent
\underbar{References}

\ref G. Garc\'\i a-Calder\'on, J.L. Mateos and M. Moshinsky, Phys. Rev. Lett.
{\bf 74}, 337 (1995).
\ref G. Garc\'\i a-Calder\'on, J.L. Mateos and M. Moshinsky, Ann. of Physics
{\bf 249}, 430 (1996).
\ref J.G. Muga, V. Delgado and R.F. Snider, Phys. Rev. {\bf B52} 16381 (1995).
\ref R.M. Cavalcanti, Phys. Rev. Lett. {\bf 80}, 4353 (1998).
\ref G.Garc\'\i a-Calder\'on, J.L.Mateos and M.Moshinsky, Phys. Rev. Lett.
{\bf 80}, 4354 (1998).
\ref J. Rauch, Comm. Math. Phys. {\bf 61}, 149 (1978).
\ref A. Jensen and T. Kato, Duke Math. J. {\bf 46}, 583 (1979).
\ref M. Murata, J. Funct. Anal. {\bf 49}, 10 (1982).

\bye